\documentclass[aps,preprint,floatfix,nofootinbib,showpacs]{revtex4-1}
\pdfoutput=1
\usepackage{graphicx,color,float, amsmath}
\usepackage{hyperref}
\usepackage{MnSymbol}%
\usepackage{wasysym}%
\usepackage{color}
\usepackage{slashed}

\begin{document}

{\small
\begin{flushright}
\end{flushright} }

\title{Implications of Gamma Ray Burst GRB221009A for Extra Dimensions}
\vspace*{1cm}

\author{
Janus~Capellan~Aban, Chuan-Ren~Chen, Yuan-Feng~Hsieh, Chrisna~Setyo~Nugroho}
\affiliation{
\vspace*{.5cm}
Department of Physics, National Taiwan Normal University, Taipei 116, Taiwan 
\vspace*{1cm}}
\date{\today}

\begin{abstract}
Anomalous high energy photons, known as  GRB221009A, with  18 TeV and  251 TeV were observed by LHAASO and Carpet-2 recently. Such observation of high energy gamma-ray bursts from distant source causes a mystery since high energy photons suffer severe attenuation before reaching the earth. One possibility is the existence of axion-like particles (ALP), and high energy photons at the source can convert to these ALPs which travel intergalactically. In this paper, we study the effects of extra dimensions on the conversion probability between photon and ALPs. The conversion probability saturates and may reach almost $100\%$ for high energy photons. We show that the size of extra dimension affects the energy at which saturation occurs. The observations of high-energy photons may support the possibility of smaller extra dimension.  
\end{abstract}

\maketitle

\section{Introduction}
\label{intro}

Recently ultra high energy photons known as GRB221009A have been detected~\cite{S.Dichiara} through a range of instruments located on Earth, such as Fermi GBM~\cite{P.Veres}, Carpet-2~\cite{D.Dzhappuev}, and LHAASO~\cite{Y.Huang,LHAASO:2023kyg}. Notably, the LHAASO collaboration recorded the observation of high-energy photons, exhibiting energies extending to around 18 TeV while Carpet-2 detected a 251 TeV photons shower.
%
However, photons of cosmic origin cannot traverse vast cosmological distances and reach Earth, as they undergo substantial scattering due to the cosmic radiation background, transforming into electron and positron pairs~\cite{Gould:1966pza,Fazio:1970pr,A.Nikishov}. Consequently, the observation of energetic photons from distant sources is prohibited by conventional cosmology, unless there exists non-standard new physics that permits the occurrence of this phenomenon.

To account for the abundant detection of highly energetic photons from the gamma-ray burst GRB221009A, a mechanism is required to prevent the attenuation of photon flux as it traverses huge cosmological distances. One plausible approach involves the incorporation of axion-like particle (ALP) $a$ that interacts with photons via the relationship $\dfrac{-g_{a\gamma} a}{4} F_{\mu\nu}\tilde{F}^{\mu\nu}= g_{a\gamma} a\vec{E} \cdot \vec{B}$. The ALP represents a collective term that encompasses particles similar to the Peccei-Quinn axion $a_{PQ}$, which is renowned for its solution of the strong CP problem~\cite{Peccei:1977hh}. Given that the host galaxy of the GRB221009A source possesses a galactic magnetic field of approximately $\textit{O}(1)\:\mu$G, this magnetic field can induce the conversion of photons into ALP as caused by ALP-photon interaction. Subsequently, this ALP can travel unhindered straight to the Milky Way galaxy, and their conversion back into photons occurs due to the influence of our galactic magnetic field, enabling the observation of these highly energetic photons~\cite{Maiani:1986md,Raffelt:1987im,Zhang:2022zbm}.

Since the ALPs are Standard Model (SM) singlets, they may propagate to the large extra dimensions just like the gravitons and sterile neutrinos~\cite{Davoudiasl:2002fq,Esmaili:2014esa,Deffayet:2000pr}. The theory of large extra dimensions (LED)  was pioneered in~\cite{Arkani-Hamed:1998jmv,Antoniadis:1998ig} by proposing that the new scale of quantum gravity $M_F$ could be much lower compared to conventional Planck scale $M_P=10^{19}$ GeV to solve the gauge hierarchy problem between weak scale and gravity. The scale $M_F$ is related to the Planck scale $M_P$ by the fundamental equation~\cite{Antoniadis:1998ig} $\dfrac{M_{P}^{2}}{M_{F}^{\delta+2}}\simeq V_{\delta}\,$, where $V_{\delta}$ is the volume of the $\delta$ compact extra dimensions. In particular, for torus configuration, the volume of the extra dimensions is denoted by $V_\delta=(2\pi)^\delta R_1 R_2 ...R_\delta$, where $R_i$ for $i=1,2,...,\delta$ represents the radii of the extra dimensions. In this case, the axion can propagate into the bulk or $(4+\delta)$ dimensions while all the SM fields are confined in the brane or 4D flat-Minkowski spacetime. The higher dimensional bulk axion field has Fourier decomposition that produces a tower of axion Kaluza-Klein (KK) excitations or KK modes~\cite{Dienes:1999gw} as manifested in the brane after the compactification of the $\delta$ extra dimensions, and the lowest KK mode is identified as the ordinary ALP that we know in the brane. In this context, all the ALP KK modes possess a coupling strength $g_{a\gamma}$ to photons and possess the ability to undergo conversions to photons and vice versa, particularly in the presence of external magnetic fields.

To see the impact of KK axions on the highly energetic photons
observations on Earth, we assume the same mechanism of the conventional photon-axion mixing. The energetic photons from cosmic
sources are mixed with KK axions in the presence of the external magnetic field of the host galaxy. In return, some of the photons
are converted to KK axions that continue the journey unimpeded across cosmological distance. The unconverted photons are severely
attenuated due to background light by annihilating into electron-positron pairs, meaning such photons cannot travel far enough. The
KK axions will arrive at the Milky Way Galaxy and encounter galactic
magnetic fields that will be reconverted into energetic photons detectable on Earth.

The paper is structured as follows: In Section~\ref{sec:model}, we briefly discuss the model containing KK axion and photon interaction~\cite{Deffayet:2000pr,Dienes:1999gw,Chang:1999si} and derive the coupled Klein-Gordon and Maxwell equations relevant for photon to KK axion conversion. The linearization of this system of equations provides a compact expression leading to the mixing matrix $M$. After the diagonalization of $M$ in Section~\ref{sec:conversion}, we compute the conversion probabilities of photon to KK axions. The discussions of Section~\ref{sec:constraints} are about the constraints on axion parameters provided by~\cite{Galanti:2022pbg}. Here, we study the effect of axion zero mode mass $m_0$, the extra dimension radius $R$, photon-axion coupling $g_{a\gamma}$, as well the photon energy $E$ on photon to KK axions conversion. Finally, we present our conclusion in Section~\ref{sec:con}.

\section{Model}
 \label{sec:model} 
In the framework of LED, where hierarchical radii are present with a dominant radius denoted as $R$ much larger than the others, the bulk system is effectively equivalent to a 5D extra-dimensional model with extra-dimensional radius $R$, as discussed in works such as~\cite{Barbieri:2000mg}. In this paper, we focus on this type of LED model. We recall the generic bulk axion and photon dynamics considered in~\cite{Deffayet:2000pr,Dienes:1999gw,Chang:1999si}, the action is given as 
\begin{align}
\label{eq:Lagrangians}
S_4 =\int d^4x \Bigg[\sum_{n} \Bigg( -\dfrac{1}{2}\{ \partial^\mu a^{(n) } \partial_\mu a^{(n) }+ m_{n}^{2} a^{(n)^2} \} -\dfrac{ g_{a\gamma} a^{(n) } }{4} F_{\mu\nu} 
 \tilde{F}^{\mu\nu} 
 \Bigg)-\dfrac{1}{4}F_{\mu\nu}F^{\mu\nu}     \Bigg]
\end{align}
such that $m_{n}^{2}=\dfrac{n^2}{R^2}$ for $n\ge 1$ are the masses square of the nonzero mode of Kaluza-Klein (KK) axions $a^{(n)}$, while $\tilde{F}_{\mu \nu}=\frac{1}{2} \epsilon_{\mu\nu \rho \sigma} F^{\rho \sigma}$ is the dual of the electromagnetic field tensor $F_{\mu \nu}=\partial_\mu A_{\nu}-\partial_\nu A_{\mu}$, and $\epsilon_{\mu\nu \rho \sigma}$ is the antisymmetric tensor satisfying  $\epsilon_{0123}=+1$. In addition, $m_{0}$ denotes the mass of zeroth mode $a^{(0)}$ of KK axions. Here, each of the KK axion states $a^{(n) }$ couples to photon with coupling strength $g_{a\gamma}$. 

We assume that the photon field $A^\mu=(0,\vec{A})$ produced from the source, travels across its host galaxy and encounters external magnetic field $\vec{H}_{0}$. Throughout the discussion, we assume that a monochromatic beam of light is traveling in $u$ direction  with the vector component of the photon field $\vec{A}$ as in~\cite{Deffayet:2000pr}
\begin{align}
\label{vectorphoton}
\vec{A}(u,t)=i\,(\vec{A}_\times(u), \vec{A}_\plus(u), 0)e^{-i\omega t}\,,
\end{align}
where the coordinate $u$ is along the direction of propagation. The corresponding electric and magnetic fields are expressed as 
\begin{align}
\label{electricandmagnetic}
\vec{E}\equiv -\partial_t \vec{A}(u,t)= (\omega\vec{A}_\times(u), \omega\vec{A}_\plus(u), 0)e^{-i\omega t}\,, \\
\vec{B}\equiv \nabla \times \vec{A}=(-i\partial_u A_{\plus}(u), i\partial_u A_{\times}(u),0)e^{-i\omega t}\,.
\end{align}
The equations of motions for the pseudoscalar field axion and photon field that can be determined from the action in Eq.~\eqref{eq:Lagrangians} are given by the coupled Klein-Gordon and Maxwell equations~\cite{Deffayet:2000pr}
\begin{align}
\label{KG-Max1}
(\Box - m_{\vec{p}}^{2})a^{(n)} =\dfrac{g_{a\gamma}}{4}F_{\mu\nu} 
 \tilde{F}^{\mu\nu}\,, \\
\label{KG-Max2} 
 \partial_{\alpha} F^{\alpha \beta}=-g_{a\gamma}\,\partial_{\alpha}\Big[\sum_{n}a^{(n)}  \tilde{F}^{\alpha \beta} \Big]\,.
\end{align}
Next, we define a set of orthonormal basis
\begin{align}
\label{basis}
\vec{e}_{||}=\frac{\vec{u}}{u},\:\:\: \vec{e}_{\times}=\frac{\vec{H}_{0\perp}}{H_{0\perp}},\:\:\:\vec{e}_\plus \,,
\end{align}
such that $\vec{H}_{0\perp}$ is the orthogonal component of the external magnetic field $\vec{H}_{0}$ with respect to the direction of the photon propagation $\vec{u}$. Furthermore, by writing the axion field as $a^{(\vec{p})}(u,t)=a^{(\vec{p})}(u)e^{-i\omega t}$, Eqs.~\eqref{KG-Max1}~-~\eqref{KG-Max2} can be simplified into~\cite{Deffayet:2000pr}
\begin{align}
\label{KG-Max3}
(\Box - m_{n}^{2})a^{(n)} =-g_{a\gamma}\omega H_{0\perp}A_{\times}\,,\\
\label{KG-Max4}
 \Box A_{\lambda}=- g_{a\gamma} \omega H_{0\perp} \delta_{\lambda \times} \sum_{n}a^{(n)}\,,
\end{align} 
where $\lambda$  stands for photon polarization vector $\lambda=+, \times$. 
Note that when $\lambda=+$ the axion states decouple from the photon. This is because only the parallel component of the photon to $\vec{H}_{0\perp}$ contributes to the of axion-photon dynamics. Therefore, in our case, we neglect the component $A_{\plus}$ of the photon.

Before simplifying Eqs.~\eqref{KG-Max3}~-~\eqref{KG-Max4} directly, let's consider the dispersion relation of the photon and external magnetic field. It is assumed that the external magnetic field varies very much larger than the photon wavelength in space. Thus, we may write the D'lambert operator $\Box$ as
\begin{align}
\label{dispersionrelation}
\Box=(\omega^2 + \partial_{u}^{2} )=(\omega + i \partial_{u} ) (\omega - i\partial_{u})\simeq (\omega+k)  (\omega - i\partial_{u})=2\omega (\omega - i\partial_{u})\,,
\end{align}
having the refractive index $n$ satisfies the form of general dispersion relation  $\omega\simeq n k$ with  $|n-1|<<1$ in our case. 
Thus, Eqs.~\eqref{KG-Max3}~-~\eqref{KG-Max4} can be further simplified as \cite{Deffayet:2000pr}
\begin{align}
\label{KG-Max5}
(\omega-i\partial_u - \dfrac{m_{n}^{2}}{2\omega})a^{(n)} =-\dfrac{g_{a\gamma}}{2} H_{0\perp}A_{\times}\,,\\
\label{KG-Max6}
 (\omega - i\partial_{u}) A_{\lambda}=- \dfrac{g_{a\gamma}}{2}  H_{0\perp} \delta_{\lambda \times} \sum_{n}a^{(n)}\,.
\end{align} 
These coupled equations may be written in a compact form
\begin{align}
\label{lin:KG-Max}
(\omega-i\partial_u+ M)\begin{pmatrix}
A_{\times}\\
a^{({0})}\\
\vdots\\
a^{({n})}\\
\vdots
\end{pmatrix}=0\,,
\end{align} 
with the corresponding mixing matrix $M$ given by
\begin{align}
\label{mixingmatrix}
M =\begin{pmatrix}
 \Delta_{\times}& \Delta_M& \Delta_M&  \hdots& \\
\Delta_M& \Delta_a^{({0})}& 0& 0&\hdots& \\
\Delta_M& 0& \ddots& \ddots& \ddots& \\
\vdots& \vdots& \vdots&\Delta_a^{({n})}& \ddots& \\
\vdots& \vdots& \vdots& \vdots& \ddots&
\end{pmatrix}\,.
\end{align} 
Here, $\Delta_M=\dfrac{g_{a\gamma}H_{0\perp}}{2}$, $\Delta_{a}^{(n)}=-\dfrac{m_{n}^{2}}{2\omega}$, 
and $\Delta_{\times}=-\dfrac{\omega_{pl}^{2}}{2\omega}$ with $\omega_{pl}=\dfrac{4\pi\alpha n_e}{m_e}$. The latter expression is the photon mass with $n_e$, $m_e$, and $\alpha$ denoting electron density, electron mass, and the fine structure constant, respectively. 

\section{Photon-Axion Conversion Probability}
\label{sec:conversion}
The high-energy photons detected on Earth can be attributed to two potential astronomical sources: magnetars, which are newly born, rapidly rotating stars with intense magnetic fields, and supernovae, resulting from the collapse of exceptionally massive stars~\cite{Granot:2015xba,MacFadyen:1998vz}. In this scenario, we consider that the extremely energetic photons 
may
originate from the mentioned astronomical sources and travel as concentrated gamma-ray jets 
throughout the universe. Concerning the mechanism of the conversion $\gamma \rightarrow a \rightarrow \gamma$~\cite{Wang:2023okw}, these photons pass through the host galaxy, they interact with magnetic fields leading to the conversion of a portion of the photon fluxes into KK axions. The resulting KK axions remain undisturbed as they journey into the intergalactic region, while the photons are scattered by the background photons present in the surrounding space. The KK axions are then converted back into photons as it arrives in the Milky Way galaxy. 

To compute the conversion probability, we need to solve Eq.~\eqref{lin:KG-Max} with the procedure provided in~\cite{Raffelt:1987im} by considering the solution of the form
\begin{align}
\label{solution}
\vec{V}(u)=e^{-iMu} e^{-i\omega u} \vec{V}(0)\,,
\end{align}
where $\vec{V}(u)$ is given by a column vector
\begin{align}
\label{columnsolution}
\vec{V}(u)=
\begin{pmatrix}
A(u) \\
a^{0}(u)\\
a^{1}(u)\\
\vdots\\
a^{N}(u)\\
\end{pmatrix}\,,
\end{align}
where we use the notation $A(u)$ to be the photon state for $A_\times(u)$ and $a^{n}(u)$ are the axion Kaluza-Klein modes for $n=0,1,2,..., N$. Suppose we consider the eigenvector $\vec{v}_n=(E_{n\gamma}, E_{n0},...,E_{nN})^T$ corresponds to the eigenvalue $\lambda_n$ of $M$, where $n=\gamma,0,1,...,N$ which satisfies the eigenvalue equation $(M-\lambda_n\cdot I ) \vec{v}_n=0$ 
\begin{align}
\label{eigeneq}
\begin{pmatrix}
\Delta_{\times}-\lambda_n& \Delta_M& \Delta_M& \hdots& \Delta_M&\\
\Delta_M& \Delta_a^{({0})}-\lambda_n& 0& \hdots& 0&\\
\Delta_M& 0& \Delta_a^{({1})}-\lambda_n& \hdots& 0&  \\
\vdots& \vdots& \vdots& \ddots& 0& \\
\Delta_M& 0& 0& 0& \Delta_a^{({N})}-\lambda_n&
\end{pmatrix}
\begin{pmatrix}
E_{n\gamma}\\
E_{n0}\\
E_{n1}\\
\vdots\\
E_{nN}\\
\end{pmatrix}
=0\,.
\end{align}
This eigenvalue equation yields the $N+2$ relations 
\begin{align}
\label{relations}
E_{n\gamma} (\Delta_\times-\lambda_n) + \Delta_M \sum_{k=0}^{N} E_{nk}&=0\,,\\
\Delta_M E_{n\gamma} + (\Delta_a^{({0})}-\lambda_n) E_{n0}&=0\,,\\
\Delta_M E_{n\gamma} + (\Delta_a^{({1})}-\lambda_n) E_{n1}&=0\,,\\
&\vdots\\
\Delta_M E_{n\gamma} + (\Delta_a^{({N})}-\lambda_n) E_{nN}&=0\,.
\end{align}
More compactly, aside from the first entry in Eq.~\eqref{relations} we have the relation
\begin{align}
\label{morecompactrelaltion}
E_{nj}=\dfrac{\Delta_M E_{n\gamma}}{(\lambda_n-\Delta_a^{({j})})}\,,
\end{align} 
with $j=0,1,...,N$. Thus, the unitary matrix that diagonalizes the matrix $M$ is given by  
\begin{align}
\label{eigeneq}
U=\begin{pmatrix}
E_{\gamma\gamma}& E_{0\gamma}& E_{1\gamma}& \hdots& E_{N\gamma}&\\
E_{\gamma 0}& E_{00}& E_{10}& \hdots& E_{N 0}&\\
E_{\gamma 1}& E_{01}& E_{11}& \hdots& E_{N 1}&  \\
\vdots& \vdots& \vdots& \ddots& 0& \\
E_{\gamma N}& E_{0N}& E_{1N}& \hdots& E_{N N}&  \\
\end{pmatrix}\,.
\end{align}
From the unitarity condition $U^{\dagger} U=I$ we have 
\begin{align}
\label{unitarityconditions}
|E_{n\gamma}|^2 + |E_{n0}|^2+ \hdots + |E_{nN}|^2 = 1\,.
\end{align}
We substitute Eq.~\eqref{morecompactrelaltion} to express the mixing matrix component $E_{n\gamma}$ in terms of available parameters in $M$ and its respective eigenvalues $\lambda_{n}$, the unitarity condition yields 
\begin{align}
\label{unitarityconditions2}
&E_{n\gamma}^2 + \Delta_{M}^{2}\sum_{j=0}^{N}\frac{E_{n\gamma}^{2}}{(\lambda_n - \Delta_a^{(j)})^2}=1\,, \nonumber\\
\implies& E_{n\gamma}=\dfrac{1}{\Big(1 +\sum_{j=0}^{N}{\frac{ \Delta_{M}^{2}}{(\lambda_n - \Delta_a^{(j)})^2}} \Big)^{1/2}}\,.
\end{align}
Hence, the eigenvector of $M$ can be written as 
\begin{align}
\label{basis}
\vec{v}_n=
\begin{pmatrix}
E_{n\gamma}\\
E_{n0}\\
E_{n1}\\
\vdots\\
E_{nN}\\
\end{pmatrix}=
\begin{pmatrix}
E_{n\gamma}\\
\frac{\Delta_M E_{n\gamma}}{(\lambda_n-\Delta_a^{(0)})}\\
\frac{\Delta_M E_{n\gamma}}{(\lambda_n-\Delta_a^{(1)})}\\
\vdots\\
\frac{\Delta_M E_{n\gamma}}{(\lambda_n-\Delta_a^{(N)})}\\
\end{pmatrix}=
E_{n\gamma} \begin{pmatrix}
1\\
\frac{\Delta_M }{(\lambda_n-\Delta_a^{(0)})}\\
\frac{\Delta_M }{(\lambda_n-\Delta_a^{(1)})}\\
\vdots\\
\frac{\Delta_M }{(\lambda_n-\Delta_a^{(N)})}\\
\end{pmatrix}=
\dfrac{1}{\Big(1 +\sum_{j=0}^{N}{\frac{ \Delta_{M}^{2}}{(\lambda_n - \Delta_a^{(j)})^2}} \Big)^{1/2}}
 \begin{pmatrix}
1\\
\frac{\Delta_M }{(\lambda_n-\Delta_a^{(0)})}\\
\frac{\Delta_M }{(\lambda_n-\Delta_a^{(1)})}\\
\vdots\\
\frac{\Delta_M }{(\lambda_n-\Delta_a^{(N)})}\\
\end{pmatrix}\,.
\end{align}

If we assume initially that we have photon-axion state $\vec{V}(0)=(A(0), a^0 (0),a^1(0),..., a^N(0))^T$, then it can be expressed as a linear combination of our basis $\vec{v}_n$ as a consequence of diagonalization, i.e.  $\vec{V}(0)=U(\vec{v}_\gamma, \vec{v}_0,...,\vec{v}_N)^T$. In this case, the initial state is given by
\begin{align}
\label{linearcombinations}
A(0)=\sum_{k=\gamma,0}^{N} E_{k\gamma} \vec{v}_k \:\:\:\:\textrm{and}\:\:\:\:\: a^n(0)=\sum_{k=\gamma,0}^{N} E_{kn}\vec{v}_k\:\:\:\:\textrm{for}\:\:\:\:n=0,1,...,N 
\end{align}
From Eq.~\eqref{solution}, the photon state initially given by $\vec{A}(0)$ would evolve into $\vec{A}(u)$ as it traverse a distance $u$
\begin{align}
\label{solutionevolution}
\vec{A}(u)&=e^{-iMu} e^{-i\omega u} \vec{A}(0)\,,\nonumber\\
&=e^{-i\omega u} \sum_{k=\gamma,0}^{N} E_{k\gamma}e^{-i\lambda_{k} u} \vec{v}_k\,,\nonumber\\
&=e^{-i\omega u} \sum_{k=\gamma,0}^{N} e^{-i\lambda_{k} u} \dfrac{1}{\Big(1 +\sum_{j=0}^{N}{\frac{ \Delta_{M}^{2}}{(\lambda_k - \Delta_a^{(j)})^2}} \Big)} (1, \frac{\Delta_M}{(\lambda_k-\Delta_a^{(0)})}, \frac{\Delta_M}{(\lambda_k-\Delta_a^{(1)})},\cdots,\frac{\Delta_M}{(\lambda_k-\Delta_a^{(N)})} )^T\,. 
\end{align}
Thus, the conversion probability of photon to KK axions is given by~\cite{Deffayet:2000pr}
\begin{align}
\label{eq:conversion}
P(A(0)\rightarrow a(u) )&=1-P(A(0)\rightarrow A(u))\,,\nonumber\\
&=1-|\langle{A(u)}|{A(0)}\rangle|^2\,, \nonumber\\
&=1-|\langle{e^{-i\omega u} \sum_{k=\gamma,0}^{N} E_{k\gamma}e^{-i\lambda_{k} u} \vec{v}_k}|{A(0)}\rangle|^2\,, \nonumber\\
&=1-\Big|\sum_{k=\gamma,0}^{N} e^{+i\lambda_{k} u} f_{k}^{2}\Big|^2\,,
\end{align}
with the normalization factor $f^{2}_{k}$ given by
\begin{align}
\label{eq:fksquare}
f_{k}^{2}= \dfrac{1}{\Big(1 +\sum_{j=0}^{N}{\frac{ \Delta_{M}^{2}}{(\lambda_k - \Delta_a^{(j)})^2}} \Big)}\,, \text{with}\, \sum^{N}_{k} f^{2}_{k} = 1\,,\textrm{for}\:\:\:k=\gamma,0,1,...,N\,.
\end{align}
One may write the conversion probability in a more compact form~\cite{Deffayet:2000pr}
\begin{align}
\label{eq:probCompact}
P(A(0)\rightarrow a(u) ) \equiv P(\gamma \rightarrow a) = 2\, \sum^{N}_{i,j} f^{2}_{\lambda_{i}} f^{2}_{\lambda_{j}} \,\text{sin}^{2} \frac{(\lambda_{i}-\lambda_{j}) u}{2} \,\,\,,\textrm{for}\:\:\:i,j=\gamma,0,1,...,N\,.
\end{align}
We solve numerically the eigenvalues of matrix $M$ and further compute the photon to axion conversion probability given in Eq.~\eqref{eq:conversion}. 
To reduce the computer power, we include up to 10 KK axions in our numerical results, since our conclusions remain even more KK modes are involved.

\section{Results and Discussion} 
\label{sec:constraints}
In the energy range between 100 GeV and 300 TeV relevant for our study as well as the extra dimension size less than 30 $\mu$m~\cite{Lee:2020zjt},  the components of the matrix $M$ exhibit some hierarchies $\Delta_{\times} \approx 0$, $\Delta^{(n)}_{a} >> \Delta^{(0)}_{a}\,, \Delta_{M}$. Therefore, the eigenvalues of the matrix $M$ may be well approximated as
\begin{align}
\lambda_{\gamma} & \approx \frac{\Delta^{(0)}_{a}}{2} - \Delta_{M}-\frac{(\Delta^{(0)}_{a})^{2}}{8\Delta_{M}} -\frac{\Delta^{2}_{M}}{\Delta^{(n)}_{a}}\,,\\
\lambda_{0} & \approx \frac{\Delta^{(0)}_{a}}{2} + \Delta_{M}+\frac{(\Delta^{(0)}_{a})^{2}}{8\Delta_{M}}\,,\\
\lambda_{n} &\approx  \Delta^{(n)}_{a} + \frac{\Delta^{2}_{M}}{\Delta^{(n)}_{a}},~\rm{for}~n\ne 0\,.
\end{align}
Furthermore, for extra dimension size of $10 \,\mu\textrm{m}\leq R\leq 30 \mu \textrm{m}$, the conversion probability can be written analytically as
\begin{align}
\label{eq:probHighR}
P(\gamma \rightarrow a) &\approx 4 f^{2}_{\gamma}\,f^{2}_{0}\,\text{sin}^{2}\left[ \left(\Delta_{M}+\frac{(\Delta^{(0)}_{a})^{2}}{8\Delta_{M}}+\frac{\Delta^{2}_{M}}{2\Delta^{(n)}_{a}}\right) u \right]\,,
\end{align}
for a fixed $n$, where the expression of the normalization factor $f^{2}_{\gamma}$ and $f^{2}_{0}$ are given by
\begin{align}
\label{eq:fgf0approx}
f^{2}_{\gamma}  \approx \frac{1}{ 1 + \frac{\Delta^{2}_{M}}{\left(\Delta_{M}+\frac{\Delta^{(0)}_{a}}{2}\right)^{2}} + \frac{\pi^{4\,\Delta^{2}_{M}}}{90\,(\Delta^{(1)}_{a})^{2}}} \,\,\,, \, f^{2}_{0}  \approx \frac{1}{ 1 + \frac{\Delta^{2}_{M}}{\left(\Delta_{M}-\frac{\Delta^{(0)}_{a}}{2}\right)^{2}} + \frac{\pi^{4\,\Delta^{2}_{M}}}{90\,(\Delta^{(1)}_{a})^{2}}}\,.
\end{align}
We may neglect $\Delta^{(1)}_{a}$ term in Eq.\eqref{eq:fgf0approx}
since it has much larger value than other parameters. In addition,
as energy gets higher, $\Delta^{(0)}_{a}$ becomes irrelevant and the
conversion probability does not depend on $E$. In this limit, both
$f^{2}_{\gamma}$ and $f^{2}_{0}$ approach $1/2$ and the contribution
of other $f^{2}_{k}$ are negligible based on the normalization
condition in Eq.~\eqref{eq:fksquare}.

In our computation, we employ the recent constraint on the size of the extra dimension which is less than 30 $\mu$m \cite{Lee:2020zjt}. 
In addition, we follow the constraints on axion to photon coupling $g_{a\gamma}$ and the zeroth mode of axion mass $m_{0}$ from ~\cite{Galanti:2022pbg,Wang:2023okw}:
\begin{align}
\label{eq:combinedconstraints}
5.0 \times 10^{-12}\:\textrm{GeV}^{-1}\leq g_{a\gamma}\leq 2.1\times 10^{-11} \textrm{GeV}^{-1}\:\:\: \textrm{for}\:\:\:10^{-10}\:\textrm{eV}\leq m_{0}\leq 2.0\times 10^{-7} \textrm{eV}.
\end{align}
To get a better understanding of our results, we put some typical values of the matrix elements $M$  
\begin{align}
\label{conversion}
\Delta_\times &\simeq \big(-1.1 \times 10^{-10}\big) \Big( \dfrac{n_e}{10^{-3}\textrm{cm}^3}\Big)\Big(\dfrac{E}{\textrm{TeV}}\Big)^{-1}\textrm{kpc}^{-1}\,,\\
\Delta_{M}&\simeq  \big(3.1 \times 10^{-2}\big)  \Big( \dfrac{g_{a\gamma}}{2\times 10^{-11}\textrm{GeV}^{-1}}\Big)\Big(\dfrac{H_{0\perp}}{\mu\textrm{G}}\Big)\textrm{kpc}^{-1}\,,\\
\Delta_0&\simeq (-7.8\times 10^{-3}) \Big( \dfrac{m_{0}}{ 10^{-8}\textrm{eV}}\Big)^{2}\Big(\dfrac{E}{\textrm{TeV}}\Big)^{-1}\textrm{kpc}^{-1}\,,\\
\Delta_n &\simeq (-3.04 n^2\times 10^{12}) \Big( \dfrac{\mu \textrm{m}}{R}\Big)^{2}\Big(\dfrac{E}{\textrm{TeV}}\Big)^{-1}\textrm{kpc}^{-1}\,.
\end{align}
As an illustration, setting a particular set of parameters corresponds to
\begin{align}
\label{exampleparameters}
n_e= 1.0 \times 10^{-3}\textrm{cm}^{-3}\:,u=30 \textrm{kpc},\: H_{0\perp}=5\mu \textrm{G},\: g_{a\gamma}=2\times 10^{-11}\textrm{GeV}^{-1},\:R=10\mu\textrm{m}\,,
\end{align}
the corresponding matrix elements for 2 KK axions are 
\begin{align}
\label{2kkmodesmatrixM}
M=\textrm{kpc}^{-1}\begin{pmatrix}
-1.1\times 10^{-10} \Big(\dfrac{E}{\textrm{TeV}}\Big)^{-1}  & 1.55\times 10^{-1}  & 1.55\times 10^{-1}& 1.55\times 10^{-1}\\
1.55\times 10^{-1} & -7.8\times 10^{-3}\Big(\dfrac{E}{\textrm{TeV}}\Big)^{-1}    &    0&0\\
1.55\times 10^{-1} & 0 & -3.04\times 10^{10}\Big(\dfrac{E}{\textrm{TeV}}\Big)^{-1} &0\\
1.55\times 10^{-1} & 0 & 0 & -1.22\times 10^{11}\Big(\dfrac{E}{\textrm{TeV}}\Big)^{-1}\\
\end{pmatrix}\,.
\end{align}
If we set the photon energy $E=10$ TeV, the numerical values of this matrix elements become  
\begin{align}
\label{2kkmodesmatrixMwithphotonsat18TeV}
M=\textrm{kpc}^{-1}\begin{pmatrix}
-1.1\times 10^{-11}  & 1.55\times 10^{-1}  & 1.55\times 10^{-1} &1.55\times 10^{-1}\\
1.55\times 10^{-1} & -7.8\times 10^{-4}    &    0 &0\\
1.55\times 10^{-1} & 0 & -3.04\times 10^{9} &0\\
1.55\times 10^{-1} & 0 & 0 &-1.22\times 10^{10}\\
\end{pmatrix}\,.
\end{align}
Clearly, the mixing matrix $M$ is dominated by the masses of nonzero modes of KK axion which can be found in the last two diagonal entries. The eigenvalues of the $4\times4$ matrix $M$ are given by 
\begin{align}
\label{eigenvalues ofM}
\lambda_{\gamma}=-0.1546\:\textrm{kpc}^{-1},\: \lambda_0=0.1554\:\textrm{kpc}^{-1},\:\lambda_1=-3.039\times 10^{9}\:\textrm{kpc}^{-1}\:\textrm{and}\:\lambda_2=-1.216\times10^{10}\:\textrm{kpc}^{-1} 
\end{align}
with $f_{\gamma}^{2}=0.501$, $f_{0}^{2}=0.499$, $f_{1}^{2}\approx 0$ and $f_{2}^{2}\approx 0$ as calculated using Eq.~\eqref{eq:fksquare}. The corresponding conversion probability is $P(\gamma\rightarrow a)= 0.996$. 

\begin{figure}
	\centering
	\includegraphics[width=0.6\textwidth]{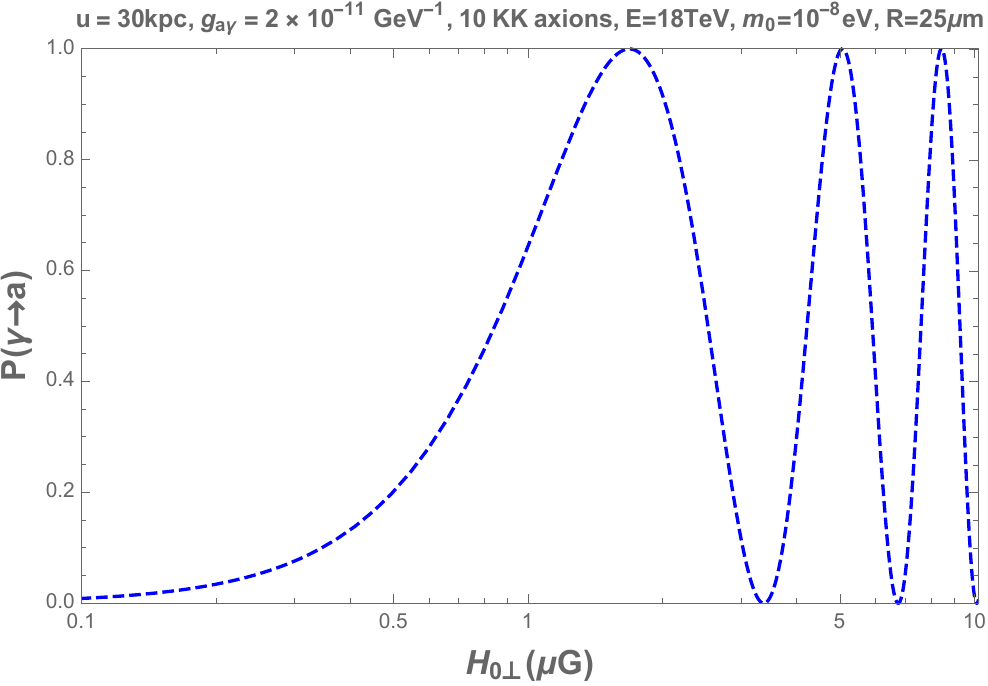}
	\includegraphics[width=0.6\textwidth]{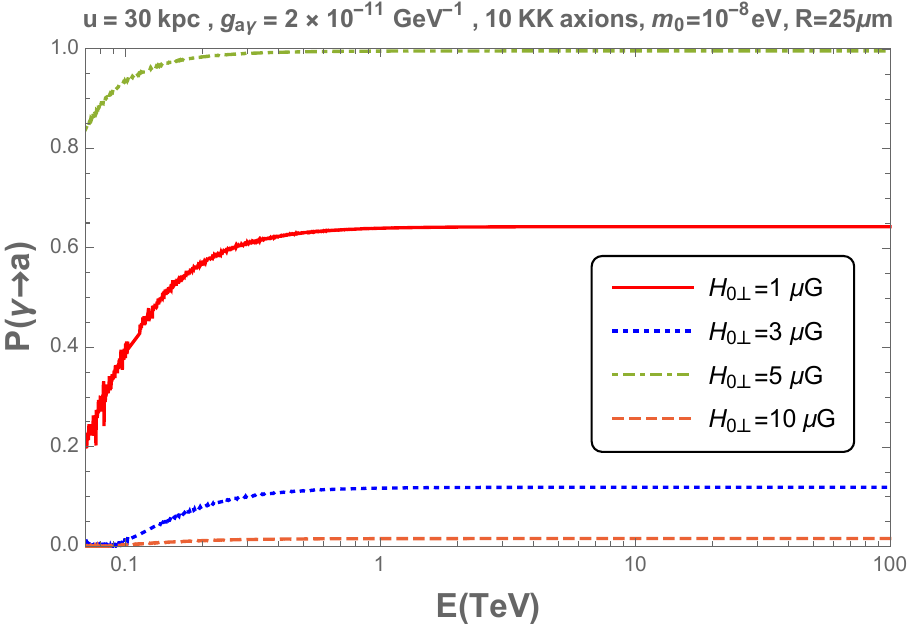}
	\caption{ The upper panel shows the conversion probability of 18 TeV photons into KK axions as a function of transverse magnetic field $H_{0\perp}$. The lower panel shows the plot of photon to KK axions conversion probability with varying external magnetic field $H_{0\perp}=1\mu$G, $3\mu$G, $5\mu$G, and $10\mu$G as a photon travels a distance of $u=30$ kpc from the source. The axion-photon  coupling strength used is $g_{a\gamma}=2\times 10^{-11}$GeV$^{-1}$ while the mass of axion zero mode is set to $m_{0}=10^{-8}$ eV. The size of the dominant extra dimension in this case is $R=25\mu$m.  }
	\label{fig:10kkaxions}
\end{figure}

In the upper panel of Fig.\ref{fig:10kkaxions}, we show the
conversion probability as a function of galactic magnetic field for photon energy $E = 18$ TeV reported by LHAASO collaboration. The
value of $P(\gamma \rightarrow a)$ oscillates as the magnetic field
changes. This behaviour is expected as one can see from Eq.\eqref{eq:probHighR} which shows that $P(\gamma \rightarrow a)$
depends on the square of sinusoidal function of $\Delta_{M} \propto g_{a\gamma}H_{0\perp} $. On the other hand, from the lower panel of
Fig.\ref{fig:10kkaxions}, $P(\gamma \rightarrow a)$ is independent
of $E$ in TeV regime since $\Delta^{(0)}_{a}$ is negligible for $m_{0} = 10^{-8}$ GeV and $R = 25\, \mu$m. In this case, both $f^{2}_{\gamma}$ and $f^{2}_{0}$ approach $1/2$ as reflected by Eq.\eqref{eq:fgf0approx}.   
   
\begin{figure}
	\centering
	\includegraphics[width=0.6\textwidth]{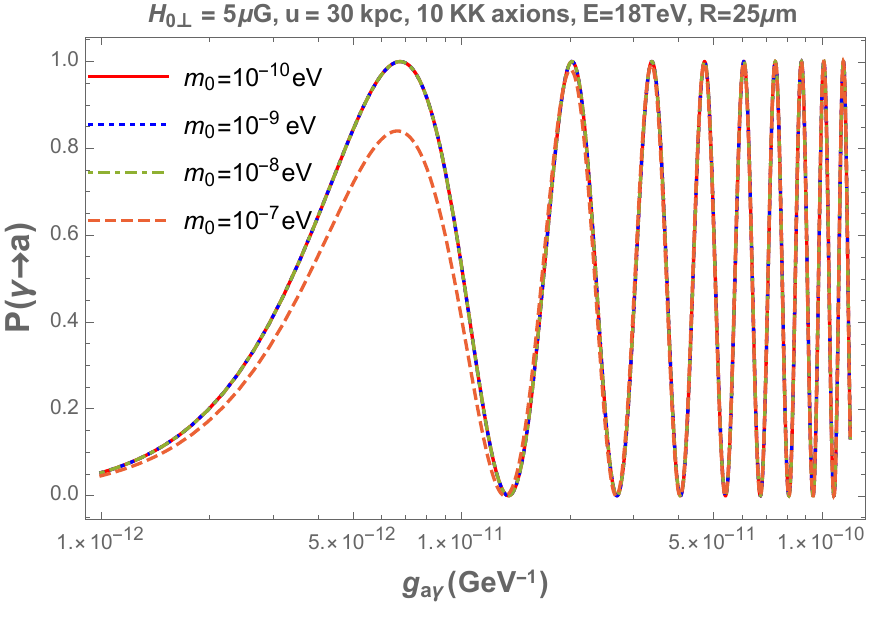}
	\includegraphics[width=0.6\textwidth]{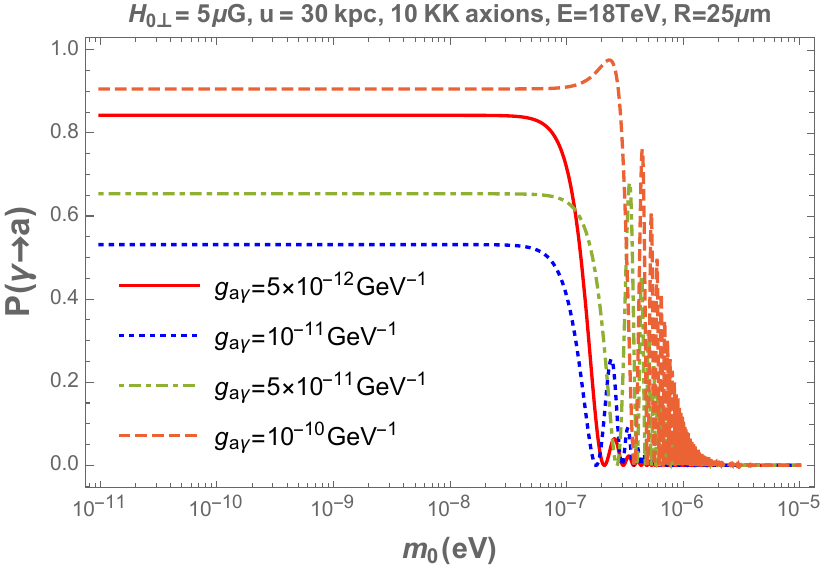}
	\caption{ The two figures show the conversion probability on 18 TeV photons in the presence of 
	external magnetic field $H_{0\perp}=5\mu$G with radius of extra dimension $R=25\mu$m as the photon travels a distance of $u=30$ kpc from the source. 
		     The upper panel shows the plot of conversion probability  as a function of KK axion-photon coupling 
		     $g_{a\gamma}$ with different axion zero mode masses $m_{0}=10^{-10}$ eV, $10^{-9}$ eV, $10^{-8}$ eV, $10^{-7}$ eV. 
The lower panel shows the plot 
		     of photon to KK axions conversion probability  as a function of axion zero mode mass 
		     $m_{0}$ with different couplings $g_{a\gamma}=5\times 10^{-12}$GeV$^{-1}, 10^{-11}
		     $GeV$^{-1},  	
		     5\times 10^{-11}$GeV$^{-1}, 10^{-10}$GeV$^{-1}$.  }
	\label{fig:varyingcouplings10kkandmasses}
\end{figure}

The oscillatory behaviour of $P(\gamma \rightarrow a)$ with respect to $g_{a\gamma}$ can also be seen in the upper panel of Fig.\ref{fig:varyingcouplings10kkandmasses}. Here, for $m_{0} > 10^{-8}$ eV at $g_{a\gamma} = 7 \times 10^{-12} \text{GeV}^{-1}$, the value of $\Delta^{(0)}_{a}$ becomes larger. Consequently, the value of $f^{2}_{0}$ is decreased ($\Delta^{(0)}_{a}$ has negative value), resulting in lower value of conversion probability, see Eq.\eqref{eq:probHighR} and Eq.\eqref{eq:fgf0approx}. As $g_{a\gamma}$ increases, the value of $\Delta_{M}$ dominates $\Delta^{(0)}_{a}$, and the oscillation amplitude returns to the situation that $f^{2}_{\gamma} = f^{2}_{0} \approx 1/2 $. On the other hand, the lower panel of the same figure shows that $P(\gamma \rightarrow a)$ drops significantly for $m_{0} > 10^{-7}$ eV.   
This is because as $m_{0}$ increases, the value of $\Delta^{(0)}_{a}$ would be very close to $\Delta_{M}$ leading to the cancellation in the denominator of $f^{2}_{\gamma}$ in Eq.\eqref{eq:fgf0approx}
\begin{align}
\label{eq:fgammaCancel}
f^{2}_{\gamma} \approx \frac{1}{1+ \Delta^{2}_{M} / \epsilon^{2}} \approx \frac{\epsilon^{2}}{\Delta^{2}_{M}}\,,
\end{align} 
where $\epsilon = \Delta_{M} - |\Delta^{(0)}_{a}| /2 $ is close to zero. As a result, the conversion probability is suppressed while retaining its oscillatory behaviour controlled by $\text{sin}^{2} ((\Delta^{0}_{a})^{2} \, u/8\Delta_{M})$.  
\begin{figure}
	\centering
	\includegraphics[width=0.45\textwidth]{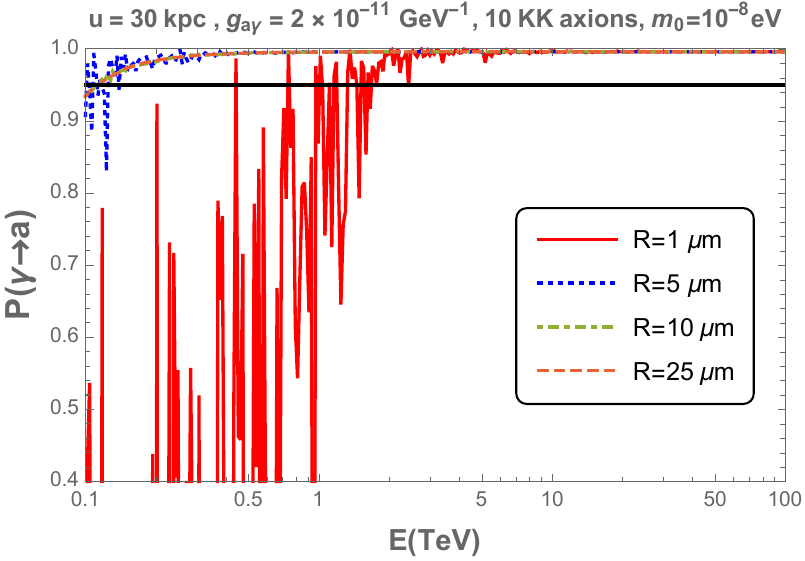}\includegraphics[width=0.45\textwidth]{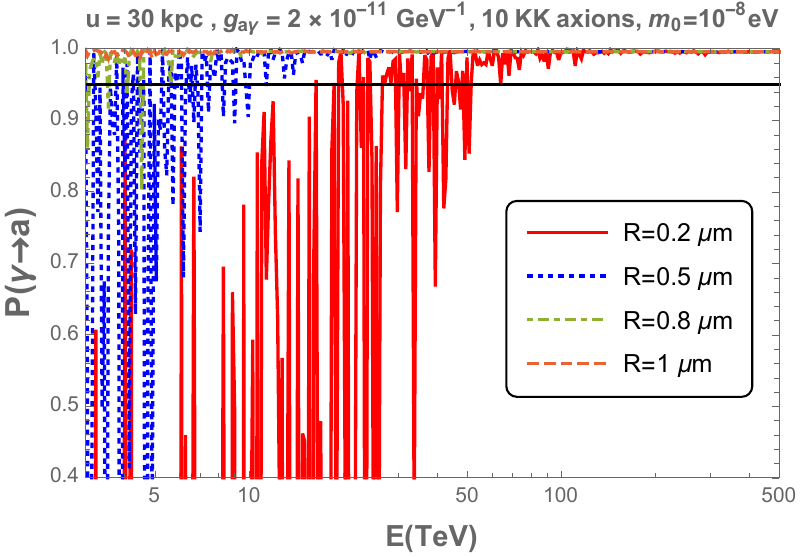}
	\includegraphics[width=0.45\textwidth]{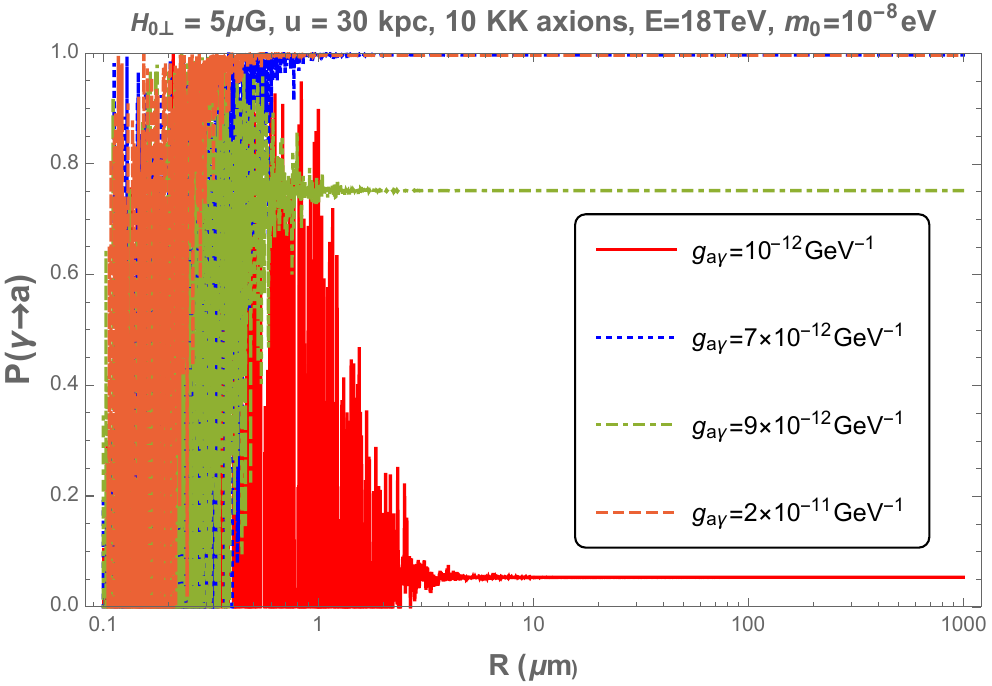}
	\caption{The upper plots show the conversion probability of photon into KK axions with different sizes of extra dimensions. The black horizontal line indicates the 95 percent conversion probability. The lower plot shows the conversion probability of 18 TeV photons into KK axions as a function of the size of extra dimension $R$. The magnetic field we consider is $H_{0\perp}=5\mu$G with axion zero mode mass $m_0=10^{-8}$eV.}
	\label{fig:differentR}
\end{figure}

Next, the effects of $R$ on the conversion probability is displayed
in Fig.\ref{fig:differentR}. At both low as well as high energy
regime, $P(\gamma \rightarrow a)$ oscillates rapidly when $R$ is less
than 10 $\mu$m as one can see from the upper panel. The same
behaviour is also observed in the lower panel of Fig.\ref{fig:differentR} where $P(\gamma \rightarrow a)$ saturates at $R\geq$ 10 $\mu$m. Here, one can not use Eq.\eqref{eq:probHighR} to describe conversion probability as it is suppressed by $\Delta^{2}_{M}/2\Delta^{(a)}_{n}$. The rapid oscillation is due to the
difference between two eigenvalues $\lambda_{m} - \lambda_{n} \approx \Delta^{(a)}_{m} - \Delta^{(a)}_{n}$
appearing in the sinusoidal argument of Eq.\eqref{eq:probCompact} with $m \neq n$ and $m,n \neq \gamma, 0$. Thus, the corresponding conversion probability is dominated by KK higher modes
\begin{align}
\label{eq:ProbKK}
P(\gamma \rightarrow a) = 2\, \sum^{N}_{m,n} f^{2}_{m} f^{2}_{n} \,\text{sin}^{2} \frac{(\Delta^{(a)}_{m} - \Delta^{(a)}_{n}) u}{2} \, = 2\, \sum^{N}_{m,n} f^{2}_{m} f^{2}_{n} \,\text{sin}^{2} \frac{(m^{2} - n^{2}) u}{4 E\, R^{2}}\,,
\end{align}
where $m,n = 1,...,N$ and $R \leq 10 \mu$m. We can see that the conversion probability is given by the
superposition of highly oscillated function leading to spiky
oscillation pattern. Such behaviour is very different from the
smooth oscillation given by Eq.\eqref{eq:probHighR} which is
controlled by the photon and zero mode axion.  
As an additional remark, the amplitude of conversion probability $f^{2}_{n}$ in this case reads
\begin{align}
\label{eq:ProbRapid}
f^{2}_{n} = \frac{1}{1+\Delta^{2}_{M}\, \left(\frac{1}{(\lambda_{n}-\Delta^{(1)}_{a})^{2}}+...+\frac{1}{(\lambda_{n}-\Delta^{(n)}_{a})^{2}}+...+\frac{1}{(\lambda_{n}-\Delta^{(N)}_{a})^{2}}\right)}\,.
\end{align}
Note that the term $1/(\lambda_{n}-\Delta^{(j)}_{a})$ in the
denominator of $f^{2}_{n}$ is maximized at resonance for $j=n$.
The value of $f^{2}_{n}$ and the associated conversion probability at resonance are given by
\begin{align}
\label{eq:fRes}
f^{2}_{n, res} &\approx \Delta^{2}_{M}/(\Delta^{(n)}_{a})^{2} = \frac{4 E^{2}\,R^{4}\,\Delta^{2}_{M}}{n^{4}}\,,\nonumber\\
P_{res}(\gamma \rightarrow a) &= 8\sum^{N}_{m } f^{2}_{m} \, E^{2} \frac{R^{4}}{n^4}\,\Delta^{2}_{M}\,\text{sin}^{2} \frac{(m^{2} - n^{2}) u}{4 E\, R^{2}}\,, \text{for fix}\,\, n\,.
\end{align}

\begin{figure}
	\centering
	\includegraphics[width=1\textwidth]{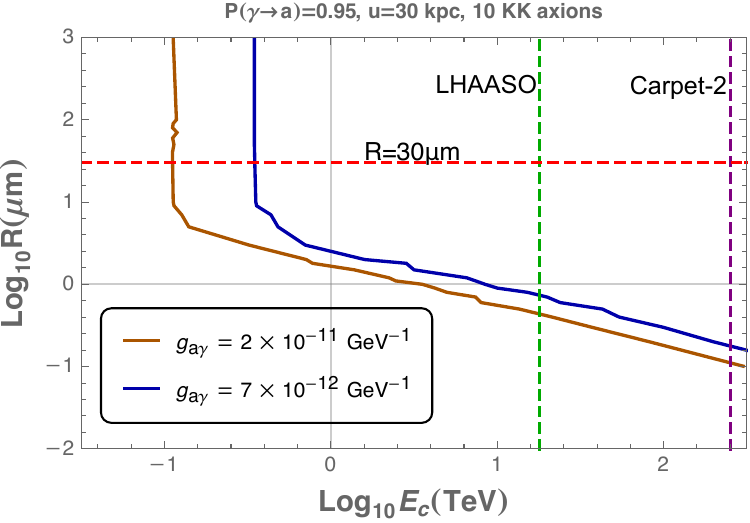}
	\caption{The plot shows the relationship between critical energy $E_c$ and extra dimension radius $R$ at 95 percent conversion probability of photons into KK axions with magnetic field $H_{0\perp}=5\mu$G and mass of axion zero mode $m_0=10^{-8}$eV. The vertical green dashed line is the maximum energy (18 TeV) detected by LHAASO~\cite{Y.Huang,LHAASO:2023kyg} while the vertical purple dashed line corresponds to 251 TeV as observed by Carpet-2~\cite{D.Dzhappuev} caused by GRB221009A. The horizontal red dashed line is the new upper bound $R=30\mu$m of the size of extra dimension from the test of Newtonian gravity at short distance~\cite{Lee:2020zjt}.}
	\label{fig:contourEcvsR}
\end{figure}

We see that all conversion probabilities converge to a constant value at high
energy, see Eq.\eqref{eq:probHighR}, the second line of Eq.\eqref{eq:fRes}, and lower panel of  Fig.~\ref{fig:10kkaxions} and Fig.~\ref{fig:differentR}. Although for the latter case, this saturation depends on the
size of extra dimension $R$. For similar $R$, the saturation occurs at higher energy. 
To optimize the high flux of axions from the source, we define the critical energy $E_{c}$
as the energy where $P(\gamma \rightarrow a ) \geq 95 \%$ for a given radius of extra dimension $R$. 
The $E_{c}-R$ plane is depicted in Fig.\ref{fig:contourEcvsR} using $g_{a\gamma} = 2\times 10^{-11} \text{GeV}^{-1}$ and $g_{a\gamma} = 7\times 10^{-12} \text{GeV}^{-1}$. These two couplings correspond to the first two peaks in the upper panel of Fig.\ref{fig:varyingcouplings10kkandmasses}. Notice that when $R$ is as large as $10~\mu\rm{m}$, the $E_c$ becomes insensitive to $R$, since the conversion probability $P(\gamma\to a)$ is dominated by the conversion between $\gamma$ and zeroth mode of axion. The red dashed line denotes
the upper limit of $R$ from~\cite{Lee:2020zjt}. The energy photons observed by LHAASO ($18$ TeV) and Carpet-2 ($251$ TeV) are shown in vertical green and purple lines, respectively.  
For $g_{a\gamma}=2\times 10^{-11}$ GeV$^{-1}$ ($ 7\times 10^{-12} \text{GeV}^{-1}$), the LHAASO data implies that, very likely, the extra dimension radius $R$ could be as small as  $0.43~\mu\rm{m}$ ($0.75~\mu\rm{m}$), while Carpet-2 supports  $R \gtrsim0.11~\mu\rm{m}$ ($0.19~\mu\rm{m}$).


\section{Conclusion }
\label{sec:con}
In this study, we have demonstrated that the detection of high-energy photons can be explained by KK axions and photon dynamics in the presence of external magnetic fields. Highly energetic photons, emitted from astronomical sources such as supernovae and magnetars, traverse the host galaxy while encountering its magnetic field. Consequently, some of these photons undergo conversion into axions, seamlessly continuing their journey across cosmological distances unimpeded. The KK axions then arrive in the Milky Way galaxy and are converted back into photons the moment they encounter the magnetic field in Milky Way, subsequently observed on Earth. Throughout our analysis, we used 10 nonzero modes of the Kaluza-Klein (KK) axions. We found that an external magnetic field of $5\mu$G is highly efficient in converting highly energetic photons into KK axions traversing a distance of $u=30$ kpc. For instance, if the photon's energy is around 18 TeV (the highest energy detected by the LHAASO collaborations) the conversion probability to KK axions reaches almost 1 if $g_{a\gamma}=7\times10^{-12}$ GeV$^{-1}$ and $g_{a\gamma}=2\times 10^{-11}$ GeV$^{-1}$ in the axion zero mode mass regime $m_{0}\leq 10^{-7}$ eV, which explains well why such high energy photon events can be observed.
Additionally, we found that the size of the extra dimension cannot be arbitrarily small, as it would diminish the conversion probability of photons to axions, hindering the detection of high-energy photons on Earth. In fact, for $H_{0\perp}=5\,\mu$G and $m_{0}=10^{-8}$ eV, achieving efficient conversion of 18 TeV photons into KK axions at a 95 percent probability or higher implies an extra-dimensional size window of $R\in[0.43\mu m, 30\mu m]$ and $R\in[0.75\mu m, 30\mu m]$ for coupling strengths $g_{a\gamma}=2\times 10^{-11}$ GeV$^{-1}$ and $g_{a\gamma}=7\times 10^{-12}$ GeV$^{-1}$, respectively. Similarly, for 251 TeV photons (the energy of photons detected by Carpet-2), the corresponding windows should be likely $R\in[0.11\mu m, 30\mu m]$ and $R\in[0.19\mu m, 30\mu m]$, respectively.

\section*{Acknowledgment}  
We would like to acknowledge the support of National Center for Theoretical Sciences (NCTS). This work was supported in part by the National Science and Technology Council (NSTC) of Taiwan under Grant No. NSTC 111-2112-M-003-006, NSTC 112-2112-M-003-007, and NSTC 112-2811-M-003-004-.



\end{document}